\documentclass[aps,prl,preprint]{revtex4}
\usepackage{epsfig}

\begin{document}

\title{ON THE GENEALOGY OF POPULATIONS:

TREES, BRANCHES AND OFFSPRINGS}

\author{Maurizio Serva}
\address{Dipartimento di Matematica,
Universit\`a dell'Aquila,
I-67010 L'Aquila, Italy}

\bigskip

\date{\today}

\begin{abstract}  

We consider a neutral haploid population whose generations are not 
overlapping and whose size is large and constantly of $N$ individuals.
Any generation is replaced by a new one and any individual has a single 
parent. 
We do not choose the stochastic rule which assigns the number of 
offsprings to any individual since results 
do not depend on the details of the dynamics, and,
as a consequence, the model is parameter free.
The genealogical tree is very complex, and distances
between individuals (number of generations from
the common ancestor)
are distributed according to
probability density which remains random in the thermodynamic
limit (large population). We give a theoretical and numerical description 
of this distribution and we also consider the dynamical aspects of the problem
describing the time evolution of the
maximum and mean distances in a single population.

\noindent
Pacs: 87.23.Kg, 05.40.-a
\end{abstract}

\maketitle
 
\section{1. Introduction}

In a population with asexual reproduction any individual has a single
parent in previous generation. 
If the size of population is constant, 
some of the individuals may have the same parent 
and, therefore, the number of ancestors of present population 
decreases if one goes backward in time.
At a finite past time one has complete coalescence 
and all population has a single ancestor.
The genealogical distance between two individuals
is simply the number of generations form the common ancestor.
The resulting genealogical tree is very complex and has many branches,
nevertheless, one would expect that in the limit
of infinite population size, some quantities would reach
some thermodynamic deterministic value. 
For example this could be the case for
the frequency of genealogical distances in a single population
or, at least, for the mean genealogical distance
obtained considering all pairs of individuals.  
On the contrary, the frequency of distances 
in a single population is random even in the thermodynamic limit.
This means that this frequency is different for different populations
and also the mean distance obtained considering
all pairs in a single population is random. 
This non self-averaging behavior is known since pioneering works 
of Derrida, Bessis and Peliti~\cite{DB,Derrida}.
In this paper we consider only the genealogical aspects of the problem,
since mutation, at this level, is only   
a measure of genealogical distance through Hamming distance.

Let us define the model.
We consider a population with asexual reproduction
and whose generations are not overlapping in time.
Any generation is replaced by a new one and any individual has a single 
parent. The size of the population
is large and constantly of $N$ individuals,
therefore, the average number of offspring of any individual is one.
The stochastic rules which assign the number of 
offsprings to any individual can be chosen 
in many ways. In fact, results 
do not depend on the details of this rule,
the only requirement is that the probability of having the same
parent for two individuals must be of order $1/N$ for large $N$.
As a consequence of the
freedom in the choice of the rule, the model is parameter free.
This is a typical situation if reproduction involves a  
fraction of order $N$ of the population.    
To be more clear we make two examples of stochastic 
dynamics which satisfy this assumption.
First rule: at any generation one half of the individuals 
(chosen at random) has no offsprings and the remaining part has two 
(see~\cite{Zhang}). With this rule the probability of having the same 
parent is $1/(N-1)$. 
The second rule (Wright-Fisher) is that any individual in the new 
generation chooses one parent at random in the previous one, 
independently on the choice of the others.
In this case the probability of having the same parent for two individuals
is exactly $1/N$.

In this paper we obtain  analytical and numerical results.
For numerical results we simulate a population of some hundred of 
individuals for $10^7$ generations according to Wright-Fisher rule.
The population is large enough to avoid finite size corrections and
time is sufficiently long to profit of ergodicity for substituting
sample averages with time averages.
Notice that we will use the world 'mean' intending mean 
over different individuals of the same population and we 
will use 'average' to intend average on many realization of the 
population process
or, equivalently, by ergodicity, average on the same population 
at different times.

The relevant quantity is the random probability 
density (rpd) of pair distances in a single population.
This quantity differs for different populations and 
changes in time for a single population.
The aim of the paper is to obtain the statistics of this density
and to obtain some informations about its dynamics.
Section 2, 3, 4 and 5 are devoted to the
first part of this program while section 6 is devoted to its
dynamical aspects.

In the final section we point out the open problems and we discuss 
the possible relevance of results for the genealogy 
of mithocondrial DNA (mtDNA) populations.

\section{2. distribution of pair distance}

The genealogical tree of a population of $N$ individuals is
determined by considering the set of all genetical distances
between them. 
The distance between two given individuals
is the number of generations from the common ancestor 
and since there are $N(N-1)/2$ possible pairs we have to specify 
$N(N-1)/2$ distances.

For large $N$ distances are proportional to $N$ so it is useful
to re-scale them dividing by $N$. 
Equivalently we can say that distances are defined as the time
from the common ancestor and contemporary define time as the 
number of generations divided by $N$.

Let us call $d(\alpha,\beta)$ the rescaled distance
between individuals $\alpha$ and $\beta$ in the population.
By definition if  $\alpha$ and $\beta$ coincide the distance 
vanishes ($d(\alpha,\alpha)=0$). 
On the contrary, for two distinct individuals $\alpha$ and $\beta$
in the same generation one has
 
\begin{equation}
d(\alpha,\beta)= d(g(\alpha),g(\beta)) +\frac{1}{N}\;\;,
\label{dynamics}
\end{equation}
where $g(\alpha)$ and $g(\beta)$ are the two parent individuals
which coincide with probability $1/N$
and are distinct individuals $\alpha'$ and $\beta'$ with probability 
$(N-1)/N$.
In other words, $d(\alpha,\beta)= 1/N$ with probability $1/N$
and $d(\alpha,\beta)= d(\alpha', \beta') +1/N$  with 
probability $(N-1)/N$.

The above equation entirely defines the dynamics of
the population, and simply state that the
rescaled distance in the new generation increases by $1/N$
with respect to the parents distance.
This dynamics can be easily simulated and at a given time 
(much larger than $N$ in order to forget initial conditions) it can be 
stopped.
The distances obtained are different for different pairs 
and their frequency can be calculated.
For finite $N$ frequency is simply the number
of pairs in a given population with given  distance $x$
divided by the total number $N(N-1)/2$ of possible pairs.

This frequency inside a single population of 500
individuals  can be seen in Fig. 1. It is immediate to observe that 
this frequency is quite wild,
due to the fact that individuals naturally cluster in subpopulation.
In fact, most of the distances assume a few of values 
corresponding to the distances between the major subpopulations.

One could think that this singular behavior would disappear in the 
thermodynamic limit of large $N$. 
On the contrary, not only the singularity remains, but
one easily realizes that this frequency remains random, 
being different for different populations and different for the 
same population at different times.
Indeed, even the mean distance in a population and the largest 
distance in a populations are random quantities in the thermodynamic 
limit as we will see in the next section. 

Let us stress again, that we use hereafter 'mean' intending mean 
over different pairs of the same population and we use 'average'
to intend average on many realization of the population process
or, equivalently by  ergodicity, average on the same population 
at different times.  
Average will be indicated by $<\cdot>$.

In spite of the frequency we can consider
the density $q(x)$ 

\begin{equation}
q(x) = \frac{2}{N(N-1)}
\sum_{\alpha>\beta} \delta (x-d(\alpha,\beta)) \,\, ,
\label{density}
\end{equation}
were the $\delta$ indicates the Dirac delta function.

This quantity is simply related to the frequency since
$q(x)\, dx$ is the number
of pairs whose distance lies in the interval 
$[x-\frac{dx}{2},x+\frac{dx}{2}]$ divided by the total number 
$N(N-1)/2$.

The random and singular nature of the density 
remains in the $N \to \infty$ limit and it is much the same 
of that of the overlap function in mean field spin glasses.
In fact, both show similar non self-averaging properties.
Indeed, the complete specification of the static 
properties of the model
would be reached if one could be able to give the probability 
distribution of  $q(x)$.   
We postpone this goal to section 5 and 
we only compute in this section the average of the 
density $<$$q(x)$$>$  and, in next two sections, 
the distribution of the largest 
distance (the distribution of the maximum
of the support of $q(x)$ )   and the first two moments of the
distribution of the mean distance.

\begin{figure}
\vspace{.2in}
\centerline{\psfig{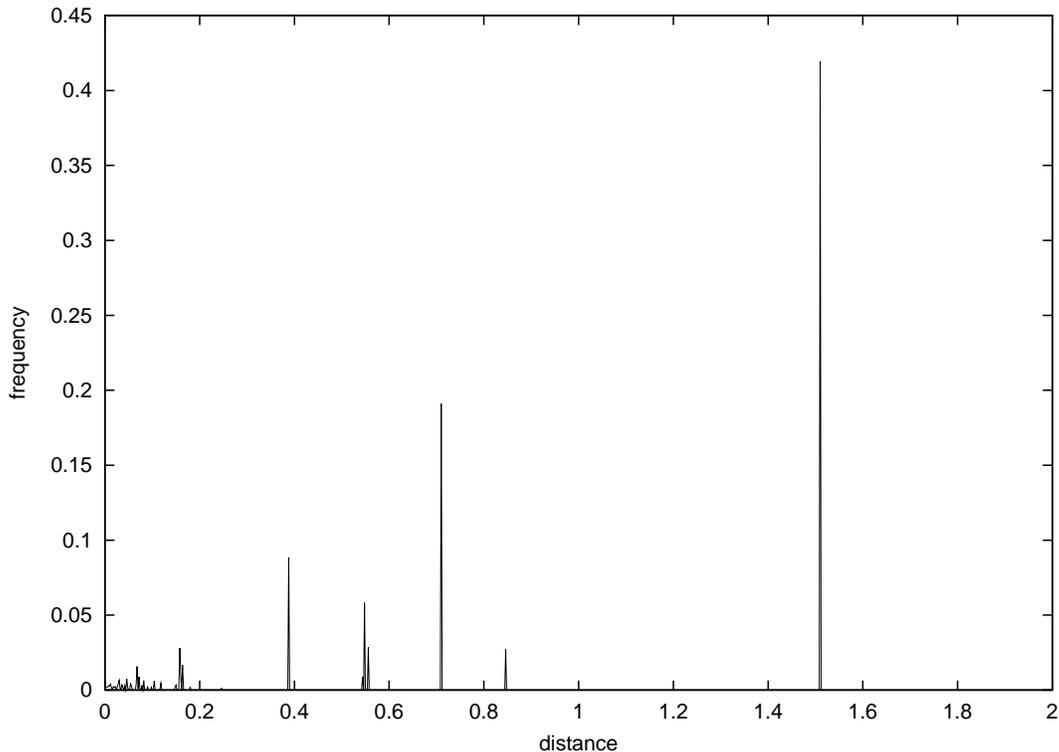}}
\bigskip
\caption{
Frequency $q(x)\, dx$  of distances in a single population.
Here we compute this quantity for a population of $500$ individuals.
Most of the distances assume a few of values 
corresponding to the distances between major subpopulations.
Notice that this frequency  is very 
different from its average $\exp(-x) \; dx$.}

\label{f1}
\end{figure}

Let us now derive the average density $<$$q(x)$$>$.
By using equation (\ref{dynamics}) and taking the average one has

\begin{equation}
<\exp(i\lambda \, d(\alpha,\beta))>\; =\;
\frac{N-1}{N} \exp(\frac{i\lambda}{N})
<\exp( i\lambda \, d(\alpha',\beta'))>
\, + \, \frac{1}{N} \,\,  .
\label{firstorder}
\end{equation} 
Since the two expectation at the left and right side of the
above equation (\ref{firstorder}) are equal, terms of order $1$ disappear
and only terms of order $1/N$ must be retained.
One gets

\begin{equation}
<\exp(i\lambda \, d(\alpha,\beta))> =\frac{1}{1-i\lambda} \,\, .
\label{fourier}
\end{equation}

This result, which holds for large $N$, implies by Fourier
inversion, that  the average probability density
for $d(\alpha,\beta)$ (i.e. $<$$ \delta (x-d(\alpha,\beta))$$>$)
is simply $\exp(-x)$.
We remark that this is not the density of 
the distances inside a single large population but the average 
distribution of two individual distance sampled over many 
stochastically equivalent populations or, which is the same,
sampled over the same population at many different times.

Notice that this result was already implicitly found in \cite{Derrida}.
In fact, in \cite{Derrida}, the genetic overlap $o(\alpha,\beta)$
of two individuals is deterministically associated to the genealogical 
distance by $o(\alpha,\beta)=  \exp(-d(\alpha,\beta)/\lambda)$
and the probability density for $o(\alpha,\beta)$ is given
as $\lambda x^{\lambda-1}$ which is directly obtainable from the 
density $\exp(-x)$ for the distance. 
The deterministic relation between distance and overlap 
is due to the infinite genome limit and
$\lambda$ is simply the inverse of the mutation rate.
Let us mention that the Hamming distance is 
linearly associated to the overlap by $1- o(\alpha,\beta)$.
In conclusion, one can easily understand
that all the complex behavior of the genetic of the populations is
due to the complexity of the structure of the genealogical tree,
the role of mutation being simply accounted by the relations 
$o(\alpha,\beta)= \exp(-d(\alpha,\beta)/\lambda)$.

We are finally able to compute the average density
using $<$$ \delta (x-d(\alpha,\beta))$$> = \exp(-x)$.
In fact, it is immediate

\begin{equation}
<q(x)> = \frac{2}{N(N-1)}
\sum_{\alpha>\beta} <\delta (x-d(\alpha,\beta))>
 = \exp(-x) \,\, .
\label{avdensity}
\end{equation}

This smooth average density is completely different from
a typical sample. To appreciate this fact is useful to
look again at Fig. 1 were  the frequency $q(x) \, dx $ is plotted.
The most important consequences of this randomness will be 
discussed in the next section.

\section{3. distribution of mean and maximum distances}

Let us introduce now two quantities which 
sinthetically describe the ``thermodynamic'' state of a population.

The first is the mean distance  

\begin{equation}
d= \frac{2}{N(N-1)}\sum_{\alpha>\beta} d(\alpha,\beta) \;\;,
\label{meandistance}
\end{equation}
which is simply the mean on a single population (and at a given time)
of the internal distances considering all the $N(N-1)/2$
possible pairs.  The above equation can be simply rewritten as
$d= \int y \, q(y) \, dy$. Since the probability density
$q(y)$ is random we expect that $d$ is also random.

The second quantity is the maximum distance  

\begin{equation}
d_{max} \; = \; \max_{\{\alpha,\beta\}} \, d(\alpha,\beta) \;\;,
\label{maximum distance}
\end{equation}
which is the largest distance in a single population, i.e. 
the maximum of the support of $q(y)$.
Again, as a consequence of the randomness of the density $q(y)$
we expect that $d_{max}$ is also random.
This quantity can be interpreted as the time from the 
common ancestor of the whole population
and it has an evident relevance in paleontology. 
In fact, mtDNA of a single species is only
transmitted by female and, therefore, can be considered as 
an haploid population. For what concerns Homo Sapiens,
$d_{max}$ is the time from the celebrated mithocondrial Eve. 

This two quantities can be studied in the context
of the coalescence problem which has been widely investigated in a number
of papers in the last two 
decades~\cite{Ald,Don,King1,King2,King3,King4,Mohle1,Mohle2,Tavare,Wat},
and is still investigated in present
times~\cite{Avi,Fill,Goh,Serva2}.
We will come back to this approach in next two sections.

Both the distances $d$ and $d_{max}$ are random quantities even in the
infinite population size limit and our goal  
is to find their density distributions $\rho(x)=<$$\delta(x-d)$$>$
and $\rho_{max}(x)=<$$\delta(x-d_{max})$$>$.
We have computed them numerically, iterating the 
dynamics (\ref{dynamics}) for $10^7$ generations for a population of 
$N=100$ individuals.
The results are shown in Fig. 2. where both the numerical densities are plotted. 

The theoretical $\rho_{max}(x)$ will be 
obtained in next section for a thermodynamic ($N=\infty$)
population and it is also plotted in Fig 1.
Coincidence between numerical and
theoretical density proves that $N=100$ can be already
considered large.

On the contrary, we have not been able to 
deduce theoretically the density $\rho(x)$.
Nevertheless, we compute its two first moments
and we show how it can be done in principle and with a lot of
work for higher moments.
First notice that from (\ref{meandistance}) one has
$<$$d$$>=<$$d(\alpha,\beta)$$>=1$.
Also notice that in the thermodynamic limit, 
again from (\ref{meandistance}), one has  
$<$$d^2$$>=<$$d(\alpha,\beta)\,d(\gamma,\delta) $$>$
where $\alpha,\beta,\gamma$ and $\delta$ are all distinct.
In fact, terms in which two or more individuals coincide are negligible
since they give a contribution of order $1/N$ to $<$$d^2$$>$.

\begin{figure}
\vspace{.2in}
\centerline{\psfig{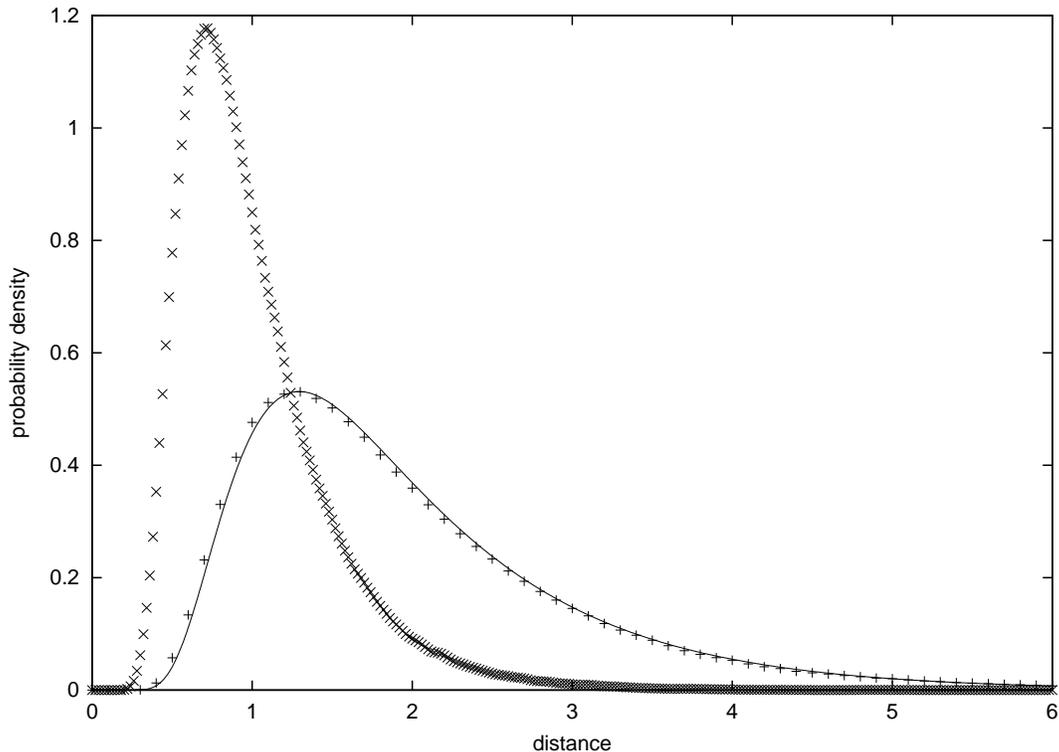}}
\bigskip
\caption{
Probability density for the maximum distance (+) 
and for the mean distance ($\times$).
The two densities are computed using a sample resulting 
from the dynamics of a population of $100$ individuals 
for $10^7$ generations.
The full line corresponds to the theoretical probability density
for the maximum distance ($N=\infty$).
Since the empirical density corresponds to $100$ individuals 
only, we deduce that a population size of $100$ is
sufficiently large to destroy finite size effects.}
\label{f2}
\end{figure}

Then, we can use again equation (\ref{dynamics}) in order to compute
the quantities $<$$d(\alpha,\beta)d(\beta,\gamma)$$>$.
To reach this goal one simply has to take into account
that any of the pairs 
which can be formed by two of the four individuals
$\alpha$, $\beta$, $\gamma$ and $\delta$  may have coinciding parents 
with probability of order $1/N$. The probability that
more then two parents coincide is of higher order and can be neglected.
Then, with the same procedure which lead to (\ref{fourier}),
(terms of order 1 disappear and only those of order 
$1/N$ are retained) one finds

\begin{equation}  
6 <d(\alpha,\beta)d(\gamma,\delta)>=
4<d(\alpha,\beta)d(\beta,\gamma)>+
2<d(\alpha,\beta)> \,\, .
\end{equation}

Again  equation (\ref{dynamics}) can be used in order to compute
the quantity $<d(\alpha,\beta)d(\beta,\gamma)>$ and obtain
from terms of order $1/N$

\begin{equation}
3<d(\alpha,\beta)d(\beta,\gamma)>=<d^2(\alpha,\beta)>
+2<d(\alpha,\beta)> \,\, .
\end{equation}

Finally, from (\ref{fourier}) not only one has
$<$$d(\alpha,\beta)$$>=1$
but also $<$$d^2(\alpha,\beta)$$>=2$.

Solving this simple system of equations one gets 
$<$$d(\alpha,\beta)d(\beta,\gamma)$$>$=4/3 and
$<$$d(\alpha,\beta)d(\gamma,\delta)$$>=11/9$
which implies
$<$$d^2$$>$=$<$$d(\alpha,\beta)d(\gamma,\delta)$$>=11/9$.
Summarizing:

\begin{equation}
<d>= 1, \; \; \; \; \; <d^2>= \frac{11}{9} \;\;,
\label{moments}
\end{equation}
which coincide with the numerical values obtained from the
string of $10^7$ generations.
The above results are related to those in \cite{Derrida}
where analogous quantities are computed for the mean overlap 
of a population.

Let us finally mention, that higher moments can be computed
using the same strategy. The result can be always found
by solving a system of linear equations.
The problem is that the number of equations in the system 
grows exponentially with the power of the moment.

\section{4. Coalescence}

The content of this section is devoted to the most studied problem
for this model: the coalescent.
The idea is very simple and goes back to the papers
of J. F. C. Kingman~\cite{King1,King2,King3,King4}
and some results has been also independently discovered 
in~\cite{DB,Derrida}.

Consider a sample of $n$ individuals in a population of size $N$.
The probability that they all have different parents in the 
previous generation is $\prod_{k=0}^{n-1}\, 
(1-\frac{k}{N})$.
Therefore, the probability that their ancestors are still 
all different in 
a past time $t$ corresponding to $tN$ generations
is $[\prod_{k=0}^{n-1}\, 
(1-\frac{k}{N})]^{tN}$. 
If $N$ is large compared to $n$ 
this quantity is approximately $\exp(-c_n t)$
where  $c_n = \frac{n\,(n-1)}{2}$. Therefore, 
the average probability density for first coalescence is 

\begin{equation}
p_n(t)=c_n \exp(-c_n t) \,\, .
\label{exponential}
\end{equation}

This expression is the probability density
for the first past time at which the ancestors 
of the $n$ individuals reduce to $n-1$.
In particular, for $n=2$ one also re-obtain the probability density
$\exp(-x)$ for the distance of a pair 
of individuals already found in section 2.

At the random time $\tau_n$ distributed according to
the exponential of parameter $c_n$, the number of ancestor 
is $n-1$ and one has to go back an exponentially
distributed  time $\tau_{n-1}$ 
of parameter $c_{n-1}$ before further coalescence and so on.
Therefore, the joint probability density 
$\prod_{k=m+1}^{n} \, p_k(t_k)$
gives the statistics for successive coalescence times 
$\tau_n \, , \tau_{n-1} \, , ..., \tau_{m+1}$
until the number of ancestor reduces to $m$.
This is the core of the celebrated coalescent,
which is mostly associated to the name of the probabilist 
J.F.G. Kingman.

If one wants to know the density distribution of time
for $n$ individuals to coalescence to $m$ ancestor
one simply has to compute the convolution
of the $n-m$ successive exponentials.
In other words this random time is simply the sum
$\sum_{k=m+1}^{n} \, \tau_k$.

The computation of the time $d_{max}$ necessary for the ancestors 
of all individuals $N$ of a population to reduce to $m$ 
needs some care in dealing with limits since, in this case, 
$n=N$. 
Nevertheless, for large $N$, one easily obtains   
$d_{max} = \sum_{k=m+1}^{\infty}\tau_k$.

In order to compute 
explicitly the statistics for
$d_{max}$ let us define $\rho_{n}(x)$
as the time density distribution of time for complete coalescence
of $n$ individuals to a single ancestor.
We have the convolution

\begin{equation}
\rho_{n}(x) = \; \int_0^x  \,
dt \, p_n(x)\;\rho_{n-1}(t-x) \,\, ,
\label{convolution}
\end{equation}
with the obvious $\rho_{2}(x)= p_2(x)= \exp(-x)$.
Then, the density for $d_{max}$ is simply 

\begin{equation}
\rho_{max}(x)= \lim_{n \to \infty }\rho_{n}(x) \,\, .
\label{coalescent}
\end{equation}

In Appendix 1 we compute explicitly the convolution
(\ref{convolution})
and we obtain the simple sum representation for
the coalescent probability density $\rho_{n}(x)$:

\begin{equation}
\rho_{n}(x)=\sum_{l=2}^{n}\;
\,(-1)^l \,(2l-1)\, c_l \;
 \left( \prod_{s=1}^{l-1} \frac{n-s}{n+s} \right)
\exp\{-\;c_l \; x\} \,\, .
\label{density}
\end{equation}

In the limit $n \to \infty$ one obtains the density for $d_{max}$

\begin{equation}
\rho_{max}(x)=\sum_{l=2}^{\infty}\;
\,(-1)^l \,(2l-1)\, c_l \;
\exp\{-\;c_l \; x\} \,\, .
\label{density}
\end{equation}
This theoretical density
(see also~\cite{DB,Derrida} and very 
recently~\cite{Fill,Goh}) is plotted in Fig 2. where it is 
compared with the density obtained by the simulation of a population 
of 100 individuals.
As already mentioned, the fact that they coincide 
so precisely can be considered 
further evidence that $N=100$ is sufficiently
large that finite size effect are negligible.

Notice that this result is far from being complete,
since it gives the distribution of the maximum distance $d_{max}$ of 
the support of $q(x)$ but it does not give more general informations
on the distribution of the density $q(x)$ itself. 
This problem will be faced in next section.

Before ending this section we would like to complete the 
description of the coalescent process by considering 
the number of offsprings of any of the ancestors.
Suppose that the total number of individuals in present generation is $N$,
then, any of the $m$ ancestors has, in present generation, 
a number $\eta_{i}^{m} \, N$ of offsprings (with $i=1,2,...,m$ 
and with $\sum_{i=1}^{m}\eta_{i}^{m}=1$).
Successive coalescence reduces the number of ancestors to $m-1$ 
and one has that any of them has
$\eta_{i}^{m-1} \, N$ offsprings in present population
($\sum_{i=1}^{m-1}\eta_{i}^{m-1}=1$).
These last numbers are such that $m-2$ of them are the same 
of the $\eta_{i}^{m}$ and one is the sum of the two remaining
of the $\eta_{i}^{k}$, correspondingly to the pair which have matched 
in a single ancestor.
This rule can be iterated until the number of ancestors reduces to a single one.

Therefore, the coalescent picture is completed by 
considering this random rule which
permits to obtain the $\eta_{1}^{m-1},....\eta_{m-1}^{m-1}$
from the $\eta_{1}^{n},....\eta_{m}^{n}$ for any $m < n$.
The rule being simply that at any step two of the numbers are chosen at 
random and summed, while the others are left unchanged.
Notice that this part of the coalescent process is independent
form the random times $\tau_1, \, \tau_2,\, ...\tau_n$.

\section{5. statistics of the random density}

We have seen that the 
time one has to go backward in order that the ancestors 
of all $N$ individuals of a population reduces to $m$ is 
$q_{m+1} = \sum_{k=m+1}^{\infty}\tau_l$.
In this case, any of the $m$ individuals will be the
ancestor of a number $\eta_{i}^{m} \, N$ of individuals
($i=1,2,...,m$)  with  $\sum_{i=1}^{m}\eta_{i}^{m}=1$.
In other words,
any of the $m$ ancestor will be at the basis of a branch
with a number $\eta_{i}^{m} \, N$ of final offsprings.
Successive coalescence reduces the number of ancestors to $m-1$,
which means the branches of two of the $m$ ancestors
are now sub-branches of a single one. 

We can now easily see how distances are distributed in a population.
At a past time $q_2$ we have that the last two
common ancestors, any of them with a number of final offsprings 
$\eta_1^2$ and $\eta_2^2$ match in a single ancestor.
Therefore, there are $(\eta_1^2 \, N\,) \, (\eta_2^2 \, N)$
pairs whose distance is $q_2$ which means that
the fraction of pairs whose distance is $q_2$ is
$p_2 = 2 \eta_1^2 \, \eta_2^2 $
according to the fact that the total number of
pairs is $N(N-1)/2$.

At a past time $q_3$ we have that two of the last three
common ancestors match in a single ancestor.
Their final offsprings before matching are
$\eta_1^3$, $\eta_2^3$ and $\eta_3^3$.
One of these three numbers equals $\eta_1^2$ or $\eta_2^2$
and the sum of the other two (say $\eta_i^3$ and $\eta_j^3$) 
equals the remaining one of $\eta_1^2$ and $\eta_2^2$.
The fraction of pairs whose distance is $q_3$ is
$p_3= 2 \eta_i^3 \, \eta_j^3 $.

Then we go on and at time $q_m$ we have that two of the last $m$
common ancestors match in a single one.
The numbers of their offsprings before matching are 
$\eta_1^m, \,  \eta_2^m, \, ....  \,\eta_m^m$.
There are $m-2$ of these numbers
which equal $m-2$ of the  
$\eta_1^{m-1} , \, \eta_2^{m-1}2, \,...., \, \eta_{m-1}^{m-1}2$
and two (say $\eta_i^m$ and $\eta_j^m$) whose
sum equals the remaining one.
The fraction of pairs whose distance is $q_m$ is 
$p_3= 2 \eta_i^m \, \eta_j^m$.

It is now quite clear how the probability density
$q(x)$ looks like.
First, its support is only in the random times $q_k$
with $2 \le k < \infty $ and 
where  $q_k = \sum_{l=k}^{\infty}\tau_l$.
Second, the fraction of pairs corresponding
to distances $q_1 , q_2...$  is    
 $p_1, p_2,...$ which satisfy $\sum_{k=1}^{\infty}p_k=1$.

Therefore, the probability density $q(x)$ is

\begin{equation}
q(x) = \sum_{l=2}^{\infty} p_l\delta(x-q_l) \,\, .
\label{rdensity}
\end{equation}

Now, what we need is to give the statistics of the numbers  $q_2,q_3,...$
and  $p_2,p_3,...$.

The first part of this program is simple. In fact, since
the probability for the sequence $\tau_2,\tau_3,...$
is   $\prod_{k=2}^{\infty} \, c_k \, \exp(-\, c_k \, t_k)$
and since $q_{l+1}=q_l-\tau_l$ we have that joint probability for 
the sequence $q_2,q_3,...$ is

\begin{equation}
\prod_{k=2}^{\infty} \, 
c_k \, \exp [- \, (k-1) \, q_k ] \,\, ,
\label{jdensity}
\end{equation}
where it is assumed that $q_k \ge q_{k+1}$.

The second part of the program is a little more difficult.
First we stress that $\tau_2,\tau_3,...$ are independent from the sizes 
$\eta_i^j$  and, therefore,  
the the random sequence $p_2,p_3,...$ is independent from 
the sequence $q_2, q_3,...$ .

To obtain the statistics for $p_2,p_3,...$ we have to
consider the coalescence rule for the $\eta_i^j$ described
at the end of previous section.
According to it, one has the conditional probability 
$p (\eta_{1}^{m-1}, ....,\eta_{m-1}^{m-1}   \, |
 \,\eta_{1}^{m}, ....,\eta_{m}^{m})$  which 
is constant whenever the rule is satisfied and vanishes elsewhere.

Assume that the limit of infinite $N$ holds,
in this case the numbers $\eta_{1}^{m},\, ....\, \eta_{m}^{m}$
may assume any real value on  $\sum_{i=1}^{m} \, \eta_{i}^{m} = 1$. 
 Also assume that the probability density 
$q (\eta_{1}^{m}, ....,\eta_{m}^{m})$ 
is constant on $\sum_{i=1}^{m} \eta_{i}^{m} =1$
and vanishing elsewhere.
Than, the probability density for
the $q( \eta_{1}^{m-1}, ....,\eta_{m-1}^{m-1})$
is also constant on $\sum_{i=1}^{m-1}\eta_{i}^{m-1} =1$ 
and vanishing elsewhere.
This property can be easily verified using 
the above described conditional probability density (see also~\cite{King1}).

Therefore, if the probability density
$q (\eta_{1}^{n}, ....,\eta_{n}^{n})$
is constant for a given $n$ then
it is  constant for any $m \le n$ and the process rule
can be easily reversed. In other words, the conditional probability
$p (\eta_{1}^{m}, ....,\eta_{m}^{m}   \, |
 \,\eta_{1}^{m-1}, ....,\eta_{m-1}^{m-1})$ can be computed
from $p (\eta_{1}^{m-1}, ....,\eta_{m-1}^{m-1}   \, |
 \,\eta_{1}^{m}, ....,\eta_{m}^{m})$
and $q (\eta_{1}^{m}, ....,\eta_{m}^{m})$  .
The only point which need some care is 
to show that for infinite $N$ the density
$q (\eta_{1}^{n}, ....,\eta_{l}^{n})$
is, indeed, constant for a given $n$ .
This task is accomplished in Appendix 2.

Using the above results we obtain with a simple calculation that
the conditional density $p (\eta_{1}^{m}, ....,\eta_{m}^{m}   \, |
\,\eta_{1}^{m-1}, ....,\eta_{m-1}^{m-1})$ corresponds to the following
reversed rule: 
one chooses at random $1 \le i \le m-1$ 
with probability  $\eta_i^{m-1}$
and cut  $\eta_i^{m-1}$ in two segments 
 $\chi \eta_i^{m-1}$ and
 $(1-\chi)\eta_i^{m-1}$ with $\chi$ uniformly distributed
between 0 and 1.
Then one has that two of the   $\eta_{1}^{m}, ....,\eta_{m}^{m}$
are   $\chi \eta_i^{m-1}$ and  $(1-\chi)\eta_i^{m-1}$
while the others $m-2$ equals the remaining $m-2$
of the   $\eta_{1}^{m-1}, ....,\eta_{m-1}^{m-1}$.

There is a picture that is useful to 
shortly describe the rule.
Consider a square with unitary surface. 
Choose a point  $x_2$ with uniform distribution between 0 and 1.
Put it on the basis of the square, then it 
will cut the unitary segments in two parts which
can be identified with $\eta_1^2$ and  $\eta_2^2$. 
Therefore, the shaded area in Fig 3a is $p_2$.
Then choose a second point $x_3$ with uniform distribution between 
0 and 1. Put it on the basis
and it will be in one of the two previously created segments
with probability proportional to their size.
Furthermore, the cut in the chosen segment 
will be uniformly distributed on it.
Then, $p_3$ will be the darker shaded area of Fig 3b.
Then choose a third point $x_4$ with uniform distribution between 
0 and 1. Put it on the basis of the square and
it will be in one of the three previously created segments
with probability proportional to the their size.
Furthermore, the cut in the chosen segment 
will be uniformly distributed on it.
Then $p_4$ will be the darkest shaded area of Fig 3c.
Then you can go on  and the whole square
will be shaded when the operation is repeated infinite times.

In conclusion, we have the complete rule for constructing
$q(x)$ since we have the
joint probability for $q_2, \, , q_3, \, ....$ and
we have the simple rule exemplified in Fig. 3 for
the joint probability for $p_2, \, p_3, \, ....$ .

Indeed, we are not able to find explicitly this second joint probability
density and, at this stage, the result 
is little more than transforming a complicate random 
dynamics (\ref{dynamics}) in a simpler random rule
of repeated fractioning.

\begin{figure}
\vspace{.5in}
\centerline{\psfig{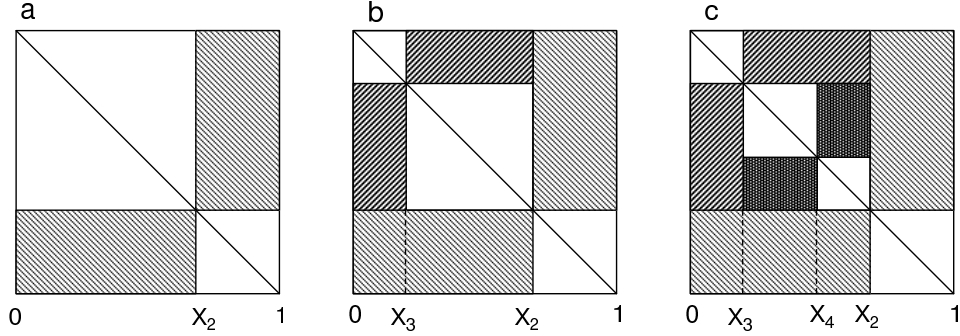}}
\caption{The point $x_2$ is chosen with uniform distribution 
on [0, 1] than the shaded area in (a) is $p_2$.
The point $x_3$ is also chosen with uniform distribution on [0, 1],
then, $p_3$ is the darker shaded area in (b).
The point $x_4$ is also chosen
with uniform distribution on [0, 1], then, 
$p_4$ will be the darkest shaded area in (c).
The whole square will be shaded when the operation is repeated 
infinite times correspondingly to the fact that 
$\sum_{i=2}^{\infty} p_i =1$.}
\label{f3}
\end{figure}

Before ending this section we would like to make some comments.
Notice that the average value of $d_{max}$ ($d_{max}$ is $q_2$)
is $<$$d_{max}$$>=2$, which means 
that a population has a common
ancestor at a past time which corresponds 
in average to $2N$ generations.
On the other side, the time for the number of ancestors
to reduce to two is $q_3$ with  $<$$q_3$$>=1$. 
Therefore,
the number of generation one has to step backward in order
that ancestors reduce to a pair 
is $N$ in average and, then, it is necessary 
to step backward $N$ more generation in average before  
ancestors reduce to a single.
This also means that for any realization of the process,
the density $q(x)$ has an isolated Dirac delta corresponding to
the maximum distance while all the remaining support is concentrated
in a segment whose size is, in average, one half of the maximum
distance. 

This means that any population naturally splits
in two subpopulation which are the descendants
of two different ancestors. 
All the distance between pair of individuals
from the two different subpopulation coincide with 
the maximum distance $d_{max}$ of average 2,
on the contrary, the distances inside the two 
subpopulations are in average smaller than 1.
This considerations will find a motivation in the final discussion.

\section{6. Dynamics}

The dynamics of the model is in principle very complicated,
since one should be able to describe the time evolution
of the density $q(x)$.
As a more reachable goal one could try to describe the time 
evolution of the maximum distance and of the mean distance in a 
population.
The behavior of these quantities is shown in Fig.4
where we plot the
maximum distance and mean distance
of the individuals of a single population as a function of time.
The two distances result from the dynamics of a 
population of $N=500$ individuals generated for
$5000$ generations
which correspond to a time lag $10$.    
Notice that both distances are subject to abrupt
negative variations due to the extinction of large
subpopulations.
In particular the maximum distance increases constantly
until has a large negative jump due to the extinction
of one of the two subpopulations
which are composed by the offsprings of 
one of the last two 
ancestors of the whole population.
At this point one of the other ancestor become 
the last common ancestor of all population and
the maximum distance is reduced consequently.  

\begin{figure}
\vspace{.2in}
\centerline{\psfig{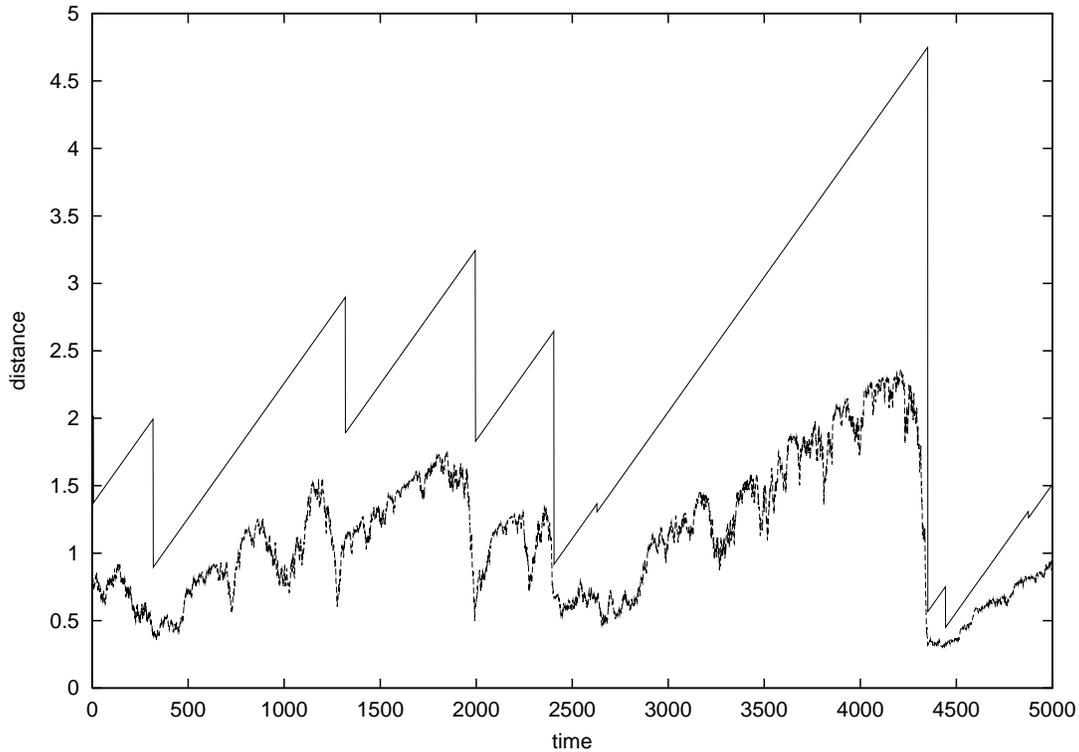}}
\bigskip
\caption{ 
Maximum distance (full line) and and mean distance
(slashed line) of the individuals of 
a single population as a function 
progressive generation number.
The two distances result from the dynamics of a 
population of $N=500$ individuals generated for
$5000$ generations (time is number of generations divided by $N$).
The size of the population is
sufficiently large to destroy finite size effects.    
Notice that both distances are subject to abrupt
negative variations due to the extinction of large
subpopulations.}
\label{f4}
\end{figure}

The full line in Fig. 4 gives the maximum distances
at all times, while the maximum distances at the time of jumps correspond 
to its relative maxima.
Furthermore, the jump sizes are
the differences between relative maxima and subsequent relative 
minima of the same full line.

How are distributed jumps and relative maxima?
In order to compute the densities of these two quantities
we have generated a dynamics for a population of $100$ individuals 
for $10^7$ generations corresponding to about of $10^5$ relative maxima.
Both densities are plotted in Fig. 5.

The probability density for the size of jumps, shown in Fig. 5, 
is compatible with $\exp(-x)$ which is quite surprising.
In fact, it is true that the density of distance between 
the last two ancestors is $\exp(-x)$, but, this is true 
in average with respect a generic time, and not
necessarily at the times of jumps.
Even more surprising is that the empirical density 
of the maximum distance at the times
of jumps coincides (Fig. 5) with the theoretical
density (\ref{density}).
The second, in fact, gives the statistics
of the maximum density at a generic time.
In other words, the first is the density
of the relative maxima of the full line in Fig. 4,
while the second is the density of all the points
of the same full line.

\begin{figure}
\vspace{.2in}
\centerline{\psfig{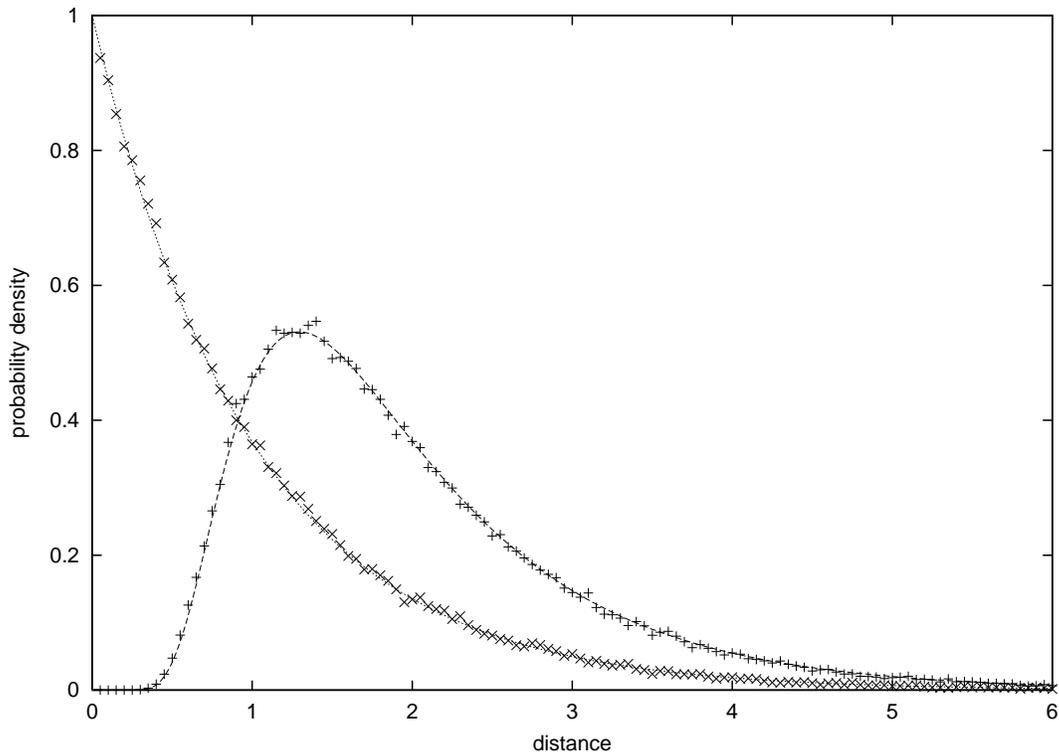}}
\bigskip
\caption{
In this picture we have the probability density of 
the maximum distance (+) at the time of jumps 
and the probability density for the size of jumps ($\times$).
The maximum distances at the time of jumps correspond to the
relative maxima of the maximum distance process (see Fig. 4).
These densities ($100$ individuals) are computed from a sample 
of $10^5$ maxima.
Comparison with the full line (the theoretical probability 
density for the maximum distance) 
shows that the statistical properties of
the maximum distance and of the maximum distance at
the times of jumps are the same.
The probability density for the size of jumps ($\times$)
is compared with $\exp(-x)$.} 
\label{f5}
\end{figure}

Finally, we would find the statistics
for the lags between jumps.
Again, in order to compute this density 
we have generated a dynamics (\ref{dynamics})
for a population of $100$ individuals 
for $10^7$ generations corresponding to about $10^5$ lags.
The result, as it can be seen in Fig. 6, is that 
lags between jumps are exponentially 
distributed according to $\exp(-x)$.

In order to understand this behavior is sufficient to consider that
the time of jumps is when one of the two subpopulation 
corresponding to the two more recent ancestors of all individual
extinguish. 
Assume that at a given time $t$ the number of the individuals belonging
to the two subpopulations is $yN$ and $(1-y)N$, then at the next 
generation (at time $t+1/N$) this numbers are  $zN$ and $(1-z)N$.
Assuming Wrigh-Fisher rule we have that the
probability density for $z$ given $y$ is

\begin{equation}
\rho(z \; | \; y )= \left( \begin{array}{cc}
N \\ Nz \end{array} \right) \;
y^{Nz} \; (1-y)^{N(1-z)} \,\, ,
\label{conditional}
\end{equation}
which in particular implies the two following conditional expectations
for $z$ and $z^2$ given $y$:

\begin{equation}
<(z \; | \; y )> = y \;\;\;\;\;\;
<(z^2 \; | \; y )> = y^2 + \frac{y\,(1-y)}{N} \,\, .
\label{expected}
\end{equation}

It is now simple to construct the diffusion limit of (\ref{conditional}).
In fact, if we write $x(t+\frac{1}{N})=z$ and $x(t)=y$ we have
$<x(t+\frac{1}{N})-x(t)>=0$ and $<(x(t+\frac{1}{N})-x(t))^2>= 
\frac{x(t)\,(1-x(t))}{N}$
which can be written in the continuous time limit as

\begin{equation}
dx(t)=\sqrt{x(t)\,(1-x(t))}\;dw(t) \,\,  ,
\label{wiener}
\end{equation}
where $w(t)$ is the Brownian motion (see also~\cite{Don}). 

All what we need now to compute the statistics of the lags 
between extinctions (which are the lags between jumps) is
to compute the statistic of the hitting times for this process
at the frontier $z=0, \; z=1$.
After the process reaches the frontier 
a new process starts at a point which is 
uniformly distributed between 0 and 1.
This choice depends on 
the known fact that the two main branches of the subpopulation 
which have survived have a size uniformly distributed. 

Indeed, the statistic is simply exponential.
In order to show this fact we have simulated the above equation
for a time sufficient to have $10^5$ extintions
(hitting times).
The resulting probability density is shown in Fig. 6.
were it is also plotted the same density as it results from 
(\ref{dynamics}).

\begin{figure}
\vspace{.2in}
\centerline{\psfig{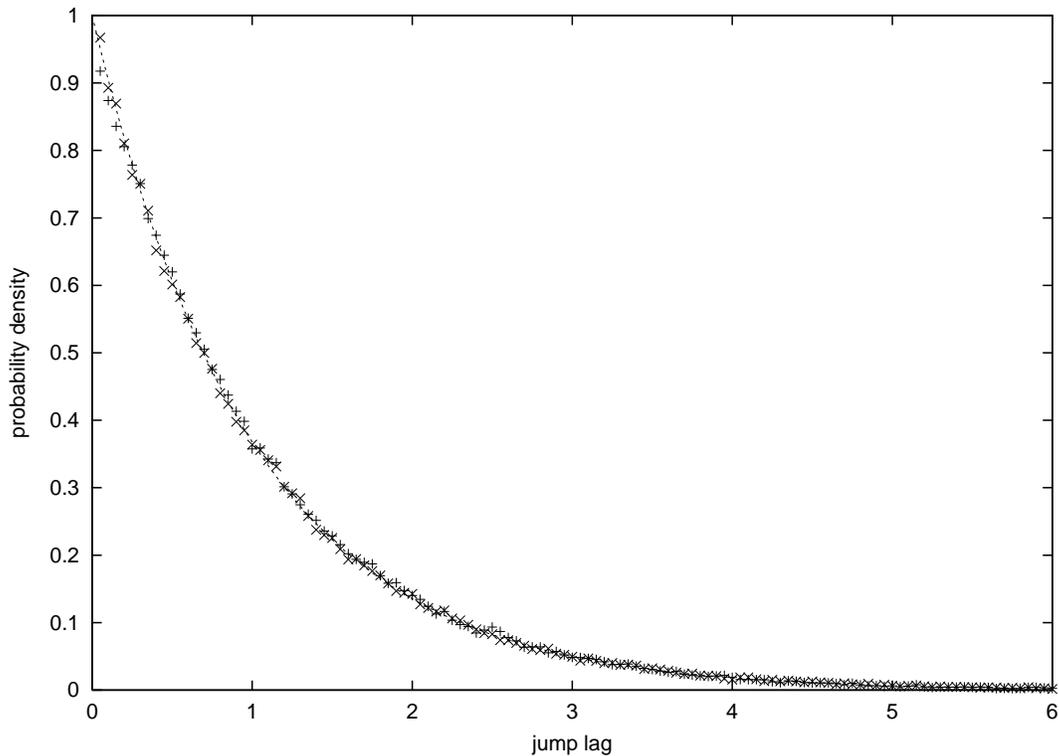}}
\bigskip
\caption{Probability density for the lag between jumps
computed from the generated dynamics of $100$ individuals ($\times$)
and probability density of lags between jumps computed 
from the simulation of the exit time for the Wiener 
process (+).
The two densities are both obtained from a sample of $10^5$
lags and both coincide with the exponential density $\exp(-x)$.}
\label{f6}
\end{figure}

\section{7. Discussion}

Before discussing the open problems concerning
the model in this paper, we would like to 
comment eventual relevance of its complex phenomenology  
for biological applications. 
Our example concerns the use of mtDna in recent 
paleoanthropological studies.  
What makes mtDNA interesting is that it is inherited only from the mother 
and it reproduces asexually at variance with nuclear DNA,
therefore, results in this paper should apply to it.
In this sense, mtDNA
of a given species which should be considers as an haploid population. 
Furthermore, assuming that mtDNA mutates at a constant rate,
the number of differences in mtDNA between two individuals
is a measure of their genealogical distance in maternal lineage. 
Let us illustrate our example.

In the years from 1997 to 2000 
some mtDNA from three different specimen of Neandertal was 
extracted~\cite{Krings1997,Krings2000} and
short strands of the hyper-variable region (HVR1 and HVR2)
were amplified using Polymerase Chain Reaction (PCR).

Two different mtDNA sequences
were extracted from the first specimen.
For the first sequence modern humans differed from each other 
in 8.0 $\pm$3.1 positions,
while the Neandertals differed in 27.0 $\pm$2.2 positions from
modern humans. 
For the second mtDNA sequence
modern humans differed from each other 
by 10.9 $\pm$5.1  and the Neandertals
differed in  35.3 $\pm$2.3 from modern humans.
The mtDNA sequence of the 
second Neandertal was compared with a 
particular modern human sequence, known
as the reference sequence. 
Difference from reference modern human sequence was in 22 position, 
(27 for the first Neandertal)
while the two Neandertals differed from each other in 12 positions.     
Sequencing of a third Neandertal mtDNA 
confirmed previous result since the difference
from modern humans was in 34.9 $\pm$ 2.4 positions.

The conclusion was that, given the
above ranges in differences, the Neandertals mtDNA 
is statistically 
different from modern humans mtDNA. 
We think that this conclusion is doubtful since 
results and discussion in section 5 show that this
situation is absolutely typical. This fact can be also
appreciated in Fig. 1.

Let us continue with our example.
A modern human fossil, 60,000 years-old,
(older then the three Neandertal fossils)
was discovered in 1974 in the 
dry bed of Lake Mungo in  Australia.
Recently, some sequences
of his mtDNA were extracted from fragments of his 
skeleton~\cite{Adcock} and from differences between   
Mungo mtDNA and living aborigines mtDNA and conclude
that Mungo man belongs to a lineage diverging 
before the most recent common ancestor of contemporary humans. 
Also in this case, the argument is doubtful, in fact,  
rapid extinctions of mtDNA subpopulations at all scales are well 
evident in Fig. 4.  

The conclusion was that
both Neandertals and Mungo man should be eliminated from our ancestry. 
It was argued, in fact,  that 
the distance of Neandertals from living humans was too large
and that Mungo carried a mtDNA which disappeared from modern humanity. 
On the contrary, it is possible that
this is not true since one can observe this mtDNA phenomenology
in a perfectly inter-breeding and (nuclear DNA) homogeneous population.
In fact, sexually reproducing nuclear DNA
has a completely different statistics~\cite{JM,DMZ,Serva} and
in large populations the distance for almost all pairs of individuals
coincide with the average value~\cite{Serva}.

We would like to conclude by a list of open problems.
First of all we would like to
compute the probability density for the mean distance $d$.  
We are in principle able to painfully
compute all moments of the random variable $d$
following calculations in section 3 but we are not able at the moment
to give an explicit expression  of its probability density.
More important, we would like to find
the explicit joint probability for $p_2, \, p_3, \, ....$.
Notice that we are able to give this probability only
indirectly by the processes in paint-boxes of Fig. 3.
Finally, we would like to characterize the time behavior
of the maximum distance, which means to find the
process for $d_{max}$ of which we have a realization in Fig. 4.

\section{Appendix 1}

We give here a very simple derivation of an 
explicit representation for $\rho_n(x)$. 
We first show that

\begin{equation}
\rho_{n}(x)= \left( \prod_{s=2}^{n} c_s \right) 
\sum_{l=2}^{n}
\left( \prod_{s=2; s \neq l}^{n} \frac{1}{c_s - c_l} \right)
\exp(-c_l x) \,\, .
\label{a1}
\end{equation}
It can be directly verified that 
the above equation holds for $n=3$ according to
(\ref{exponential}) and (\ref{convolution}).
Furthermore, 
assuming that it holds for a given $n-1$, 
from (\ref{exponential}) and (\ref{convolution})
we obtain 

\begin{equation}
\rho_{n}(x)= \left( \prod_{s=2}^{n} c_s \right) 
\sum_{l=2}^{n-1}
\left( \prod_{s=2; s \neq l}^{n} \frac{1}{c_s - c_l} \right)
\exp(-c_l x)
-
\left( \prod_{s=2}^{n} c_s \right) 
\sum_{l=2}^{n-1}
\left( \prod_{s=2; s \neq l}^{n} \frac{1}{c_s - c_l} \right)
\exp(-c_n x) \,\, .
\label{a2}
\end{equation}
If we compare (\ref{a1}) and (\ref{a2})
we see that they coincide provided

\begin{equation}
- 
\left( \prod_{s=2}^{n} c_s \right) 
\left( \prod_{s=2; s \neq n}^{n} \frac{1}{c_s - c_n} \right)
\exp(-c_n x)
-
\left( \prod_{s=2}^{n} c_s \right) 
\sum_{l=2}^{n-1}
\left( \prod_{s=2; s \neq l}^{n} \frac{1}{c_s - c_l} \right)
\exp(-c_n x) = 0 \,\, ,
\label{a3}
\end{equation}
which holds assumed that

\begin{equation}
\sum_{l=2}^{n}
\left( \prod_{s=2; s \neq l}^{n} \frac{1}{c_s - c_l} \right)
 = 0 \,\, .
\label{a4}
\end{equation}

Therefore, all what we need to prove the preliminary 
representation (\ref{a1}) is that (\ref{a4}) holds.
To reach this goal let us define the Lagrange polynomial

\begin{equation}
Q(x)=
\sum_{l=2}^{n} c_l
\left( \prod_{s=2; s \neq l}^{n} 
\frac{c_s -x}{c_s - c_l} \right) \,\, .
\label{a5}
\end{equation}
It is immediate to verify that for every $l$
such that $2 \leq l \leq n$ one has  $Q(c_l)=c_l$.
Since the degree of the polynomial is at most $n-2$ 
and since it crosses the above $n-1$ points
it is necessarily $Q(x)=x$  and in particular $Q(0)=0$
Then, since by definition

\begin{equation}
Q(0)=
\sum_{l=2}^{n} c_l
\left( \prod_{s=2; s \neq l}^{n} 
\frac{c_s}{c_s - c_l} \right)
 =  \left( \prod_{s=2}^{n} c_s \right) 
\sum_{l=2}^{n}
\left( \prod_{s=2; s \neq l}^{n} \frac{1}{c_s - c_l} \right) \,\, ,
\label{a6}
\end{equation}
equation (\ref{a4}) holds and (\ref{a1}) is demonstrated.

Furthermore, by a simple calculation one can show that

\begin{equation}
\left( \prod_{s=2}^{n} c_s \right) 
\sum_{l=2}^{n}
\left( \prod_{s=2; s \neq l}^{n} \frac{1}{c_s - c_l} \right)
= \,(-1)^l \,(2l-1)\, c_l  
\prod_{s=1}^{l-1} \frac{n-s}{n+s} \,\, ,
\label{a7}
\end{equation}
and finally we have the simple sum representation for
the coalescent density distribution

\begin{equation}
\rho_{n}(x)=\sum_{l=2}^{n}\;
\,(-1)^l \,(2l-1)\, c_l \;
 \left( \prod_{s=1}^{l-1} \frac{n-s}{n+s} \right)
\exp\{-\;c_l \; x\} \,\, .
\label{a8}
\end{equation}

\section{Appendix 2}

We show here that the probability density 
$q (\eta_{1}^{n}, ....,\eta_{l}^{n})$ is constant for a given $n$
when the limit of large $N$ is performed.

At a given time in the past, the number of ancestors
of all $N$ individuals of a population is $l$.
At an intermediate time, always in the past, 
the number of ancestors is $k \ge l$. 
This means that any of the $l$ individuals
is an ancestor of one ore more of the $k$ individuals
i.e., any of the $l$ branches has one or more sub-branches.
Let us call $ r_i^l$ this (integer)
number of sub-branches for individual $i$,
then,   $\sum_{i=1}^{l} \, r_i^l = k$ with $r_i^l \ge 1$.
Let us call $\Gamma_l$ the ensemble of 
$ r_1^l, \, .......\, r_l^l $  such that
$\sum_{i=1}^{l} \, r_i^l = k$ and $r_i^l \ge 1$.

Let us define $ f_l (r_1^l,\,...\, r_l^l)$ as the probability
for $r_1^l,\,...\, r_l^l$.
We first show that this probability
is constant on  $\Gamma_l$ for any  $l \le k$.

Assume that  $ f_{l+1} (r_1^{l+1},\,...\, r_{l+1}^{l+1})$ 
is constant on  $\Gamma_{l+1}$.
Coalescence implies that one of 
$r_1^l,\,...\, r_l^l$ equals the sum of two random chosen of
the  $r_1^{l+1},\,...\, r_{l+1}^{l+1}$ 
while the remaining $l-1$ coincides. 
Then, according to this rule,  $ f_l (r_1^{l},\,...\, r_{l}^{l})$ 
is constant on  $\Gamma_{l}$.
To have the proof, it is now sufficint to
remark that $ f_k (r_1^k,\,...\, r_k^k)$ 
is constant on  $\Gamma_k$. 
In fact, all $ r_i^k$ must equal one,
i.e. $f_k (1,\,...\, 1) =1$ while it vanishes elsewere.

Now, let us recall that the the number of offspring of the 
$k$ ancestors are $\eta_{1}^{k} \, N,\, ....,\eta_{k}^{k} \, N$
with  $\sum_{i=1}^{k}\eta_{i}^{k}=1$.
Assume now that a previous time the number of ancestors
is $n<k$ and assume that the number of sub-brances
of any of them is  $r_1^{n},\,...\, r_{n}^{n}$.
Then, the numbers $\eta_{i}^{n} $ will be obtained by the sum 
of $r_i^n$ of the $\eta_{1}^{k} \, , ....,\eta_{k}^{k}$
chosen at random. 

Now let us recall that $\sum_{i=1}^{n}\,  r_i^n =k$,
therefore, large $k$ implies that for almost all possible choices
on $\Gamma_{n}$ the $n$ numbers $r_i^n $ must be of order $k$.
We can define $r_i^n  =  \alpha_i^n \, k $
with $\sum_{i=1}^{n}\,  \alpha_i^n \, = 1$
and the numbers $ 0 \le \alpha_i^n \le 1 $ of order 1.
Furthermore, since $\sum_{i=1}^{k}\eta_{i}^{k}=1$,
we assume that almost all of the $\eta_{i}^{k} $ are of order
$1/k$.

We can now take the limit of large $k$ 
after the limit of large $N$.
Since the $\eta_{i}^{n} $ are the sum of
$r_i^n =  \alpha_i^n \, k$ 
of the $\eta_{1}^{k} \, , ....,\eta_{k}^{k}$
and since from definition  $\sum_{i=1}^{k}\eta_{i}^{k}=1$
with $\eta_{i}^{k} $ of order $1/k$, one has
$\eta_{i}^{n} =  \lim_{k \to \infty} \, r_i^n \,  / k = \alpha_i^n$.

Finally, since $ f_l (r_1^{n},\,...\, r_{n}^{n})$ 
is constant on  $\Gamma_{n}$   
one has that $q (\eta_{1}^{n}, ....,\eta_{n}^{n})$ 
is constant on $\sum_{i=1}^{n} \eta_{i}^{n} =1$.

\section{Acknowledgements}
We thank Davide Gabrielli, Michele Pasquini and Filippo Petroni
for many illuminating discussions.
We acknowledge the financial support of MIUR Universita' di L'Aquila,
Cofin 2004 n. 2004028108$\_$005.

\newpage

\end{document}